# Equality of Participation Online Versus Face to Face: Condensed Analysis of the Community Forum Deliberative Methods Demonstration

Eric Showers and Nathan Tindall, Todd Davies[⋆]

Center for the Study of Language and Information, and
Symbolic Systems Program
Stanford University, Stanford, California, USA
`{eshowers,ntindall,davies}@stanford.edu`

**Abstract.** Online deliberation may provide a more cost-effective and/or less inhibiting environment for public participation than face to face (F2F). But do online methods bias participation toward certain individuals or groups? We compare F2F versus online participation in an experiment affording within-participants and cross-modal comparisons. For English speakers required to have Internet access as a condition of participation, we find no negative effects of online modes on *equality of participation* (EoP) related to gender, age, or educational level. Asynchronous online discussion appears to improve EoP for gender relative to F2F. Data suggest a dampening effect of online environments on black participants, as well as amplification for whites. Synchronous online voice communication EoP is on par with F2F across individuals. But individual-level EoP is much lower in the online forum, and greater online forum participation predicts greater F2F participation for individuals. Measured rates of participation are compared to self-reported experiences, and other findings are discussed.

## 1 Introduction

The efficacy of face to face deliberation has been the subject of much discussion in the academic literature. According to some authors, it leads to better decision making and allows for a greater degree of public agency [6]. Others, however, claim that it is at best a waste of time, and at worst that it leads to bad decision making [25]. Beyond discussions

[⋆] *An extended version of this paper, with additional data, tables, figures, and analysis, is available at the SSRN open access repository* **(see reference [27]).** This research was supported by the Agency for Healthcare Research and Quality (AHRQ) under contract #HHSA290201000005C, through the American Institutes for Research (AIR), subcontract #641.02916. Additional funding was provided by the Vice Provost for Undergraduate Education at Stanford University. We thank AIR Staff members Coretta Mallery, Rikki Mangrum, Maureen Maurer, Manshu Yang, and Mark Rosenberg for helpful discussions and aid in obtaining the data. The views expressed in this paper are those of the authors, and have not been approved by AHRQ or AIR.

of effectiveness, many authors focus on specific elements of deliberation that might prove problematic or worth looking at — the most important for the purposes of this paper having to do with the marginalization of participants according to race or gender. Some of these hypothesized effects are quantitative and can be easily identified or tested for — that men speak more than women in deliberative sessions, for example [13], [29], and [21]. Other phenomena, such as domination or the idea that some demographic groups might be less likely to deliberate or have less influence, are harder to examine quantitatively [1], [25]. Echoes of these phenomena, however, might be found in the measurable quantity of an individual's contributions. Much of this literature deals with face to face deliberation. We apply such methods to online participation as well.

As Internet access becomes more widespread and allows more users to make their voices heard, its potential as a tool for public deliberation cannot be overlooked. There is already a substantial body of literature discussing the Internet's capacity for use in this regard, e.g. [8], [12], [24], [33]. This literature is particularly concerned with the ability to eliminate some of the inequalities present in face to face deliberation [26], and cites elements of online deliberation such as anonymity and remoteness as potential benefits. There are detractors as well, however, who cite issues with online deliberation including a potential lack of respect among participants and lack of Internet access among certain groups [2]. Another dimension that must be considered is facilitation style, which can impact the proceedings [31]. But while the literature often discusses online and face to face (F2F) deliberation in isolation, there are few sources that provide a *direct* comparison between the two [23]. We aim to provide a quantitative look at examples of both kinds of deliberation in order to highlight potential differences between the deliberative *modes* [8], and to examine the effects of other variables within both modes.

In order to compare online and F2F deliberation effectively, we will examine the Agency for Healthcare Research and Quality's Community Forum Project [4], [5], which gathered together a large number of deliberative groups using different methods (as described in the next section) — one online, two offline, and one mixed. The Community Forum is beyond the scale of any controlled deliberation experiment done previously, and it sought to bring together a representative sample of the U.S. population. It is also one of the few *multiple-method* experiments that provides quantitative data on populations recruited specifically for deliberation. Our analysis is drawn from transcripts of all F2F and synchronous online meetings during the Community Forum project, archives of all online forum discussions, and records of surveys filled out by the participants that measure their knowledge, attitudes, and experience both pre- and post-deliberation.

## 2  Community Forum Project

A five-arm randomized controlled trial was conducted between August and November 2012 by the American Institutes for Research. This Deliberative Methods Demonstration was intended to inform the Agency for Healthcare Research and Quality's research programs on public views regarding the usage of research evidence in health care decision making, and to expand the evidence base on public deliberation. The demonstration gathered empirical evidence about the *effectiveness* of deliberation, which has received minimal attention to date [8]. In the literature, effectiveness has been defined by the following parameters: (1) the quality of deliberative experience or discourse, (2) changes in participants' knowledge or attitudes about the deliberative topic, (3) changes in participants' empathy and concern for issues affecting the community at large, and (4) the impact of deliberation on decisions by the sponsoring agency.

For this Deliberative Methods Demonstration, participants were randomly assigned to one of four deliberative discussion methods, or to a reading-materials only (RMO) group. Participants were sampled from Chicago, IL, Sacramento, CA, Silver Spring, MD, and Durham, NC, where they were assigned into groups representative of the population of those areas with respect to gender, age, and ethnicity, as estimated by the U.S. Census. A total of 1,774 participants were recruited for the study, of whom 961 took part in a deliberative discussion method, and 377 were assigned to the RMO group.

The following deliberative question was posed to all participants: *Should individual patients and/or their doctors be able to make any health decisions no matter what the evidence of medical effectiveness shows, or should society ever specify some boundaries for these decisions?*

The participants were all given educational background materials to read. Those assigned to a discussion group then discussed the deliberative question in one of four distinct methods that have been advocated and used previously in prior public deliberations [5]. Additionally, some participants were only assigned reading materials. This was done to examine whether deliberation has a positive or negative impact on attitude change, and other measures of effectiveness. The main results of the study are reported elsewhere [4], [5]. The methods were: **Brief Citizens' Deliberation (BCD)** – one two-hour session per group, active facilitation; **Community Deliberation (CD)** – two in-person deliberative sessions with active facilitation (**CD-F2F**), each 2.5 hours long, separated by a week during which participants interacted through the online asynchronous Deme discussion board (**CD-Forum**) [9]; **Online Deliberative Polling®(ODP)** – four 75-minute online sessions with minimal student facilitation; **Citizens' Panel (CP)** – 2.5 days of deliberation with three active facilitators per group, moderated breakout groups, and unfacilitated open spaces; and the **Reading Materials Only (RMO)** Control Group – educational materials received via an email link, with no discussion (these data were not used in our investigation).

## 3   Research Questions and Previous Findings

The following research questions have been prioritized and answered in our analysis:

1. Do the medium (online versus F2F) and/or modality (e.g. speech versus text) have effects on *equality of participation* (EoP) across demographic groups (ethnicity, gender, education, age)?
2. Do online methods differ from F2F on individual-level EoP?
3. Do online methods differ from F2F in the effect of group size on EoP?
4. Do individuals who participate more online also participate more F2F?
5. What is the relationship between objective measures of EoP and self-reported experience?

While we were interested in broad differences between deliberative modes, of particular concern was the effect the deliberative environment had on the contributions of individuals based on their demographic. Some literature claims, for example, that women say less than men online, e.g. [10], [16], [17], [18], [30]. Does online deliberation bias contributions in favor of male participants? Other authors emphasize online divides related to race/ethnicity [20], educational level [20], and/or youth, e.g. [10], [30]. In terms of ethnicity, whites and males have been reported to say more than any other group in F2F deliberative settings as well [22], [25]. On the other hand, multiple studies of F2F deliberation have found that women speak as much as, or more than, men in these offline settings [11], [28], [32]. Are online settings different? Examining the quantitative data from the online sessions could help answer these questions.

Some literature shows that group size has an effect on F2F deliberation, and our aim was to use the vast quantity of data that the Community Forum Project collected to map that effect across its F2F and online modes. Because group size is less salient in online settings, these data provide a unique opportunity to test the hypothesis under different conditions. Finally, although the value of participation equality in group deliberation brings forth varying opinions in scholars, e.g. [14], [25], and [26], more unequal systems seem less desirable in cases such as public deliberation where a diversity of voices is a commonly agreed goal [7], [15].

## 4 Methods

The present study utilizes data generated in the AHRQ Community Forum Deliberative Methods Demonstration [4], [5], but this study was not conceived prior to the design of the Community Forum experiment. An optimal design for the present study would have an online forum-only group, allowing a more pure comparison between online asynchronous text forums and the other methods. The lack of such a condition reflects limitations in the budget and aims of the Community Forum project, but we believe that much can be learned by creatively exploring the data that *were* produced.

Each deliberative session was transcribed from audio and/or video recordings. For each contribution, the number of words it comprised was tabulated. From these data, the *frequency*, *volume*, and *average contribution length (ACL)* were calculated for each individual in the deliberative sessions. The frequency of contribution was calculated by dividing an individual's number of spoken continuous contributions by the total number of contributions spoken in the session. The volume of contributed words for an individual was calculated by dividing the total number of words that an individual spoke by the total number of words that were spoken in the session by all participants. The average contribution length was calculated by taking the total number of words that an individual spoke and dividing it by their number of contributions. Measuring frequency and volume as percentages was necessary to perform analysis across methods due to variation in deliberation duration and group size.

The following were considered independent variables, as self reported by each participant: *age*, *gender*, *education*, and *race/ethnicity* (Hispanic, Native American, Asian or Pacific Islander, Black or African American, White, Other). Education was self-reported as one of eight categories, increasing from "less than high school gradu-ate" to "more than 4-year college graduate." Individuals could indicate more than one race/ethnicity.

**Table 1. Mean Values: Demographic Data**

| Method | Groupts | Sessions | Individuals | Size Range | Avg. Size | Avg. Age | Fem-Prop | Education | Hispanic | Native | Asian | Black | White | Other |
|---|---|---|---|---|---|---|---|---|---|---|---|---|---|---|
| BCD | 24 | 1 | 309 | 9-14 | 13.0 | 46.9 | 0.55 | 5.49 | 0.13 | 0.02 | 0.03 | 0.27 | 0.60 | 0.10 |
| CD | 48 | 2 | 292 | 7-13 | 11.8 | 47.5 | 0.55 | 5.65 | 0.11 | 0.01 | 0.02 | 0.33 | 0.56 | 0.09 |
| CP | 12 | 3 | 98 | 20-28 | 24.3 | 48.5 | 0.57 | 5.39 | 0.10 | 0.00 | 0.01 | 0.47 | 0.43 | 0.10 |
| ODP | 72 | 4 | 262 | 5-12 | 9.5 | 45.6 | 0.52 | 5.87 | 0.11 | 0.01 | 0.01 | 0.25 | 0.64 | 0.11 |

Table 1 shows the number of transcript files that were scraped from each method, the range in attendance for sessions, and the demographic makeup of the people who participated.

Analysis was performed across methods, across media (online/offline), and by looking at isolated subpopulations in order to investigate the behavior of different ethnic and gender subgroups. Deliberative experience surveys were also administered. An equality factor, calculated to have a Cronbach's alpha value of 0.64 as a function of three of the questions, was found by exploratory factor analysis [5]. (See also [27].)

## 5 Results

We divide the results into three parts. The first part compares the ODP (synchronous voice) data with the three F2F methods. The second compares participants in the CD group who posted on the online forum (asynchronous text) with those who did not. And the third reports findings that speak to EoP across deliberative modes.

### 5.1 Synchronous Voice vs. Face to Face

Tables 2 through 4 show the frequency, volume, and average contribution length correlations with different independent variables across all five environments: ODP and CD-Forum (the online environments), the F2F component of CD (which we call CD-F2F), the BCD, and the CP environments. Significant negative effects for attendance (group size) were found with respect to frequency and volume across all four methods but no effects with respect to average contribution length were found to be significant. Significant positive effects for age were found across the various methods as well.

With respect to education, a positive relationship between contribution and self-reported education was found. For the ODP, CD-F2F, and CP conditions, no significant effect was found between gender and contribution frequency, volume, or average length. However, in the BCD condition, female identification had a significant negative correlation with volume ($\rho$ = -0.147, $p < 0.02$).

In the ODP condition, a participant indicating that they were white had a positive, significant correlation with all contribution metrics (frequency, $\rho$ = 0.107, $p < 0.001$; volume, $\rho$ = 0.191, $p < 0.001$; average length, $\rho$ = 0.116, $p < 0.01$), while black identification had a negative correlation with volume ($\rho$ = -0.134, p $< 0.001$). A similar trend was found in the BCD condition, where white identification had a positive, significant correlation with frequency and volume (frequency, $\rho$ = 0.190, $p < 0.001$, volume: $\rho$ =

**Table 2. Frequency Correlations**

| Mode | Method | Size | Age | Gender | Education | Hispanic | Native | Asian | Black | White | Other |
|---|---|---|---|---|---|---|---|---|---|---|---|
| Online | ODP | -0.304*** | 0.294** | 0.015 | 0.042 | -0.073 | 0.009 | -0.050 | -0.020 | 0.106** | -0.129*** |
| | CD-Forum | 0.010 | 0.106 | -0.016 | 0.115 | 0.030 | | -0.081 | -0.080 | -0.033 | 0.097 | -0.061 |
| F2F | CD | -0.222*** | 0.107* | -0.041 | 0.101* | -0.064 | 0.002 | -0.018 | 0.017 | 0.010 | -0.059 |
| | BCD | -0.150* | 0.211*** | -0.081 | 0.119 | 0.038 | 0.062 | -0.068 | -0.186** | 0.190*** | 0.000 |
| | CP | -0.134* | 0.134* | 0.012 | 0.118 | -0.128* | — | -0.062 | -0.127* | 0.166** | -0.070 |

**Table 3. Volume Correlations**

| Mode | Method | Size | Age | Gender | Education | Hispanic | Native | Asian | Black | White | Other |
|---|---|---|---|---|---|---|---|---|---|---|---|
| Online | ODP | -0.252*** | 0.203*** | -0.0156 | 0.160*** | -0.066 | 0.037 | -0.018 | -0.134*** | 0.191*** | -0.127*** |
| | CD-Forum | 0.008 | 0.112 | -0.029 | 0.145 | 0.006 | -0.072 | -0.074 | -0.055 | 0.122* | -0.068 |
| F2F | CD | -0.196*** | 0.047 | -0.055 | 0.138* | -0.045 | 0.022 | -0.032 | 0.030 | 0.0035 | -0.035 |
| | BCD | -0.140* | 0.087* | -0.147* | 0.136 | 0.041 | 0.038 | -0.065 | -0.127* | 0.147* | -0.004 |
| | CP | -0.121* | 0.023 | 0.001 | 0.185** | -0.011 | — | -0.065 | 0.031 | 0.006 | 0.014 |

**Table 4. Average Contribution Length Correlations**

| Mode | Method | Size | Age | Gender | Education | Hispanic | Native | Asian | Black | White | Other |
|---|---|---|---|---|---|---|---|---|---|---|---|
| Online | ODP | 0.060 | 0.063 | -0.056 | 0.167*** | 0.023 | 0.023 | 0.087* | -0.149*** | 0.116** | -0.032 |
| | CD-Forum | 0.048 | 0.091 | -0.056 | 0.151* | 0.002 | -0.085 | -0.075 | -0.030 | 0.094 | -0.055 |
| F2F | CD | -0.054 | -0.166*** | 0.032 | 0.092* | 0.043 | 0.175*** | -0.044 | 0.032 | -0.063 | 0.121** |
| | BCD | 0.032 | -0.161** | -0.117 | 0.156** | 0.032 | -0.099 | 0.019 | 0.084 | -0.047 | -0.033 |
| | CP | 0.075 | -0.112 | -0.083 | 0.129** | 0.179 | — | -0.043 | 0.024 | -0.132* | 0.163** |

*p < .05; **p < .01; ***p < .001*

0.147, $p < 0.02$), while answering the ethnicity question with "Black or African American" had negative contribution correlations (frequency, $\rho$ = -0.186, $p < 0.002$; volume, $\rho$ = -0.127, $p < 0.05$). For the CP condition, answering ethnicity with "White" had a positive, significant correlation with frequency ($\rho$ = 0.166, $p < 0.01$), while black identification had a negative, significant correlation with frequency ($\rho$ = -0.127, $p < 0.05$). No significant effects with respect to ethnicity were found for the CD-F2F method. Identifying as "Hispanic", "Native American", or "Asian or Pacific Islander" showed no systematic correlations with contribution measures across methods.

### 5.2  Findings Within the Citizens' Deliberation Hybrid Method

**Posters vs. nonposters.** Table 5 compares those who posted with those who did not post in the online forum of the CD method. The frequency, volume, and average length figures given there are for each group's average-member contributions in the F2F sessions of CD. Tables 6 through 8 compare the poster and nonposter groups in CD both online and F2F in terms of the demographic variables and sizes of the groups in which they were participating,

Posters and nonposters showed similar, significant effects with respect to attendance (group size). Posters showed positive effects for frequency and volume with respect to age, but nonposters showed no effect, a discrepancy from the findings of the other groups. Nonposters and posters shared significant negative correlations for average

**Table 5. Mean Values: Posters vs. Nonposters**

| Subset | Avg. Age | Fem-Prop | Education | Hispanic | Native | Asian | Black | White | Other | Frequency | Volume | Avg. Length |
|---|---|---|---|---|---|---|---|---|---|---|---|---|
| Posters | 50.0 | 0.58 | 5.74 | 0.12 | 0 | 0.01 | 0.35 | 0.56 | 0.08 | 0.10 | 0.10 | 35.8 |
| Nonposters | 46.7 | 0.52 | 5.52 | 0.10 | 0.03 | 0.03 | 0.30 | 0.55 | 0.118 | 0.084 | 0.085 | 41.1 |
| $p$-value | 0.31 | 0.15 | 0.25 | 0.51 | 0.01* | 0.07 | 0.24 | 0.83 | 0.12 | 0.01** | 0.03* | 0.01** |

**Table 6. Frequency Correlations**

| Mode | Subset | Size | Age | Gender | Education | Hispanic | Native | Asian | Black | White | Other |
|---|---|---|---|---|---|---|---|---|---|---|---|
| Online | Posters | -0.165* | 0.1134 | -0.069 | 0.136 | 0.006 | — | -0.063 | -0.104 | 0.158* | -0.034 |
| F2F | Posters | -0.234*** | 0.177** | -0.100 | 0.050 | -0.011 | — | -0.122* | -0.047 | 0.059 | 0.018 |
| F2F | Nonposters | -0.225*** | 0.003 | 0.016 | 0.165* | -0.166* | 0.030 | 0.074 | 0.122 | -0.082 | -0.138 |

**Table 7. Volume Correlations**

| Mode | Subset | Size | Age | Gender | Education | Hispanic | Native | Asian | Black | White | Other |
|---|---|---|---|---|---|---|---|---|---|---|---|
| Online | Posters | -0.138 | 0.124 | -0.080 | 0.181* | -0.026 | — | -0.061 | -0.126 | 0.187* | -0.051 |
| F2F | Posters | -0.170** | 0.119* | -0.135* | 0.094 | 0.000 | — | -0.115 | -0.030 | 0.046 | 0.034 |
| F2F | Nonposters | -0.249*** | -0.054 | 0.038 | 0.197** | -0.131 | 0.055 | 0.038 | 0.119 | -0.070 | -0.106 |

**Table 8. Average Contribution Length Correlations**

| Mode | Subset | Size | Age | Gender | Education | Hispanic | Native | Asian | Black | White | Other |
|---|---|---|---|---|---|---|---|---|---|---|---|
| Online | Posters | -0.082 | 0.0885 | -0.140 | 0.201** | -0.041 | — | -0.044 | -0.107 | 0.159* | -0.02 |
| F2F | Posters | 0.034 | -0.142* | -0.005 | 0.126* | 0.102 | — | -0.091 | 0.017 | -0.25 | 0.124* |
| F2F | Nonposters | -0.149* | -0.185** | 0.103 | 0.065 | -0.024 | 0.229 | -0.039 | 0.061 | -0.104 | 0.104 |

*$p < .05$; **$p < .01$; ***$p < .001$

contribution length with respect to age, however. Posters showed positive, significant correlations with respect to education only for average contribution length, while nonposters showed positive, significant correlations with respect to education for frequency and volume. No systematic, significant correlations were found for ethnicity or gender among the poster and nonposter groups, with the exception that female identification had a significant, negative correlation with contributed volume in the F2F session ($\rho = 0.197$, $p < 0.05$). In comparing the total F2F contributions of posters and nonposters (Table 5), posters' contributions were of significantly higher frequency ($p < 0.005$), significantly higher volume ($p < 0.03$), and (interestingly) their average contribution lengths were significantly less ($p < 0.01$).

**Face to face vs. asynchronous text (Deme forum).** In the F2F component of CD, group size had a statistically significant negative correlation with both frequency ($\rho = -0.222$, $p < 0.001$) and volume ($\rho = -0.196$, $p < 0.001$), but no significant impact on average length. These results were mirrored in the online component (frequency, $\rho = -0.165$, $p < 0.05$; volume, $\rho = -0.138$, $p < 0.08$) although the effect was weakened. (See Tables 6, 7, and 8 for this subsection.)

There were no significant age effects in the online case, but in the F2F sessions, frequency was positively correlated with age ($\rho = 0.107$, $p < 0.02$), while average length was negatively correlated ($\rho = -0.166$, $p < 0.001$). There were no significant effects for gender in either medium. Educational level was positively correlated with all contribution measures both online (frequency, $\rho = 0.136$, $p < 0.08$; volume, $\rho = 0.181$, $p < 0.02$; average length, $\rho = 0.201$, $p < 0.01$) and in the F2F sessions (frequency, $\rho = 0.100$, $p < 0.03$; volume, $\rho = 0.139$, $p < 0.01$; average length, $\rho = 0.092$, $p < 0.05$), with slightly stronger effects online. There were no significant race/ethnicity effects among the F2F participants, but white identification had a positive, significant correlation with all metrics in the online case (frequency, $\rho = 0.158$, $p < 0.05$; volume, $\rho = 0.187$, $p < 0.02$; average length, $\rho = 0.159$, $p < 0.05$).

**Face to face (posters only) vs. online forum in CD.** As shown in Tables 6 through 8, we also examined differences between the behavior of those who posted online and spoke offline in CD, in order to examine if the change in medium would impact individuals' contribution rates. Group size effects were consistent with the other methods, though the effect observed in the F2F mode (frequency, $\rho = -0.234$, $p < 0.001$; volume, $\rho = -0.169$, $p < 0.002$) was much stronger than in the asynchronous forum setting (frequency, $\rho = -0.153$, $p < 0.04$; volume, $\rho = -0.125$, $p < 0.08$).

Although no significant age effects were found in the online forum, the effect was significant across all metrics in the F2F setting, (frequency, $\rho = 0.176$, $p < 0.002$; volume, $\rho = 0.119$, $p < 0.04$; average length, $\rho = -0.143$ $p < 0.02$). In the F2F condition, education had a positive and significant effect on average contribution length ($\rho = 0.126$ $p < 0.02$), and was similar online (frequency, $\rho = 0.136$, $p < 0.08$; volume, $\rho = 0.181$, $p < 0.02$; average length, $\rho = 0.201$ $p < 0.01$). Among posters, women contributed less in the F2F setting (frequency, $\rho = -0.100$, $p < 0.08$; volume, $\rho = -0.135$, $p < 0.04$), though no significant gender effects were observed in the online setting. With respect to ethnicity, no systematic effects were observed in the F2F case. However, significant effects were observed for white identified posters on the online forum, who posted more than those who were nonwhite (frequency, $\rho = 0.158$, $p < 0.05$; volume, $\rho = 0.187$, $p < 0.02$; average length, $\rho = 0.159$ $p < 0.05$).

### 5.3 Equality of Participation Across Individuals

Although the most common application of the *Gini index* is its use as a measure of income inequality in a given nation, it also can be used as a general measure of inequality in a data set. In this context the Gini index ranges from 0, representing complete equality, to 1, representing complete inequality. Gini indices were calculated for each session, and the values analyzed for each medium, in order to investigate EoP differences across methods. The Gini index was calculated by the following formula, which fulfills the Transfer Principle of Inequality [19], where $X_i$ is the amount that person $i$ contributed and $P_i$ is the contribution rank of person $i$ such that the person who contributed most receives a rank of 1 and the person who contributed least a rank of $N$:

$$G = \frac{N+1}{N-1} - \frac{2}{N(N-1)\bar{x}} \sum_{i=1}^{n} P_i X_i$$

In comparing the synchronous voice method (ODP) against the other methods (Table 9), statistically significant differences were found for frequency between ODP and BCD ($p < 0.05$), and between ODP and the (Deme) Forum ($p < 0.05$). Significant differences for volume were found between ODP and both BCD and CD-F2F ($p < 0.001$). Additionally, significant differences for average contribution length were found between ODP and both BCD and CD-F2F ($p < 0.001$). For volume, ODP ($G = 0.439$) and CP ($G = 0.448$) reported the highest Gini indices, with the other F2F methods showing more modest, yet still fairly high, coefficients (BCD: $G = 0.351$; CD-F2F: $G = 0.368$). The Gini indices for the online forum when including all participants (both posters and nonposters) in the CD method were dramatically greater than for all the other methods, indicating, perhaps not surprisingly, that an optional online forum draws a more limited set of participants.

**Table 9.** Gini indices across methods

| Method | Frequency | Volume | Average Length |
|---:|:---:|:---:|:---:|
| BCD | 0.329 | 0.351 | 0.204 |
| CD-F2F | 0.335 | 0.368 | 0.214 |
| CP | 0.400 | 0.448 | 0.283 |
| ODP | 0.362 | 0.439 | 0.279 |
| CD-Forum | 0.754 | 0.702 | 0.556 |

Gini indices for each metric were also calculated based on the F2F contributions of the people who posted (Table 10). Among those who posted, a statistically significant difference between online ($G = 0.467$) and F2F ($G = 0.345$) media is prominent for volume ($p < 0.001$). Even among those who choose to participate in an online forum, there appears to be less EoP for volume and average length (thought not frequency).

**Table 10.** Gini indices among posters across mediums

| Method | Frequency | Volume | Average Length |
|---:|:---:|:---:|:---:|
| Forum | 0.316 | 0.467 | 0.302 |
| Face to face | 0.322 | 0.345 | 0.203 |

**Self reported experiences regarding equality of participation.** (See [27] for data.) In the post-deliberative experience survey, participants rated CD (F2F) and BCD the most equal of the methods, with CP being the least perceived equal, and ODP falling in between. An interesting comparison is with the measured Gini indices for each method (Table 9). The subjective equality factors roughly mirrored the pattern of Gini indices across the four rated methods, with BCD and CD(-F2F) scoring as the most equal on all three Gini measures and also on the subjective equality factor, CP scoring the least equal on all, and ODP scoring in the middle on all. Posters in the CD-Deme forum rated overall equality significantly lower in the CD-F2F environment than did nonposters, though posters were more satisfied that they personally said what they wanted to. We saw in Table 5 that posters exceeded the contributions of nonposters in the F2F sessions of CD by all three objective metrics, and their reported satisfactions can be reconciled with this fact. Posters appeared aware that they got more than their share of speaking in during the CD-F2F sessions, perhaps leading them to feel satisfied with their own participation but less satisfied that the process produced equal participation. In all but one of the methods (BCD), white identified participants rated the equality of the method significantly lower than did black identified participants, despite the fact that black identification predicted lower volume of participation in all but one of the method groups (see Table 14 in [27]).

**Individual-level equality of participation and group size online versus face to face.**
(See [27] for data.) The Gini index is a measure of how concentrated participation is across individuals (the higher the Gini, the more participation is dominated by a subgroup of participants). A natural question to ask is what effect the size of the group has on this measure. The Gini index rose substantially as group size increased in both the BCD and CP methods, but was unaffected by group size in the CD method. For the online environments, the relationship between Gini indices and group size was either flat or slightly negative for both ODP (synchronous voice) and CD-Deme (the asynchronous text forum). This provides some evidence that the two online methods each scale well, at least within the observed ranges (7-12 and 8-17 participants, respectively, for the ODP and CD groups). Adding more participants within these ranges does not seem to make participation more unequal across individuals in the two online methods, nor in the CD-F2F method. But adding participants does seem to reduce EoP in two of the F2F methods: BCD and CP.

## 6  Discussion

Some literature argues that women are less likely to participate online than are men, e.g. [3], [2], though women may be more likely to participate equally with men online than offline [23]. We found no significant negative effects on EoP for women across methods, with the exception of the BCD method, which favors men in volume. This conclusion deviates from the sizeable body of literature arguing that women speak less F2F [21], [7], [23], and from the claim that women are less active in online contributing [2], [23], in agreement with the idea that online environments do not adversely impact gender EoP (see [34]). One explanation for the F2F equality of contribution is that all of the F2F methods were facilitated, and there is evidence to support that facilitation eliminates the worst of the gender gap in deliberation [31]. The discrepancy in our findings lies with the BCD method, in which female identification negatively correlated with all contribution measures. One difference between BCD and the other F2F methods was that BCD used a male facilitator for half of the groups (rather than a female for all), although within the BCD method women were not significantly more inhibited under the male facilitator than under the female one. Despite this discrepancy, the results overall imply that the difference of mode (online vs. offline) is not causing the difference in and of itself. Indeed, in the CD condition we saw that female online forum posters participated equally with men, but the same women did not do so F2F. These findings agree with some other studies involving online deliberation, most notably another study in which participants deliberated on healthcare issues [24], but they are unique in being drawn from a within-group study. Previous examinations of online deliberation even when compared directly with F2F deliberation, have not used the same group that participated F2F when tracking online contributions.

In most of the methods, and most visibly online, there was a significant positive correlation between white identification and volume/frequency, and a significant negative correlation between black identification and the same measures. This was most prevalent in the online methods — ODP and the forum — where facilitation was the least present. The online and F2F environments showed relatively even participation levels across ethnicities in the three F2F environments, but noticeable differences in both ODP and the CD-Deme (Forum) setting (see Figure 3 in [27]). ODP is unique in that there did not seem to be a tradeoff for white participants between volume/frequency and ACL, and the correlation with all three measures was positive. These results are consistent with other findings for gender, in which the gender gap is eliminated via facilitation [31].

Noticing that age and education had consistent positive relationships with contribution measures for all metrics, a multiple regression model was generated in order to investigate to what extent these factors could compensate for other discrepancies, especially between ethnicities. As shown in Table 14 in [27], the racial difference in participation is reduced when we control for age and education. A gap persists for the two online environments: ODP and CD-Deme (Deme is the Forum component of CD), though not for the F2F environments, including CD-F2F*, which represents just Deme forum posters in the F2F component of CD. This merits further study to determine whether the media difference is robust.

For further analysis and discussion, see [27].

## 7  Conclusions

While there have been a variety of studies of online and offline deliberation, none have as large a pool of information to work with as the Community Forum project, and as such it provides a unique opportunity to provide quantitative analysis of the difference between the two modes on a scale that has not been seen before. While this paper does not represent an exhaustive report of all the conclusions that can be gleaned from the data about the effect of deliberative mode on EoP, some conclusions appear well-supported based on our analysis so far:

1. *Online effects on demographic groups' participation equality*. Overall, we see no consistent effects of online versus F2F participation for gender equality of participation in these data. There is evidence that some deliberation methods (e.g. the F2F BCD method) may adversely impact female participation, independently of the offline-online dimension, and that an asynchronous forum produces higher EoP across genders than F2F discussion. For ethnicity, the online versus F2F picture is less clear, but the online settings in this study do seem to have depressed black participation relative to whites' (see Fig. 3 in [27]). Online deliberation appears to reduce black and increase white participation somewhat, relative to F2F, even when controlling for age and educational level (see Table 14 in [27]). This provisional finding requires further investigation, but may reflect the reduced level of facilitation in the online conditions of the Community Forum experiment. Older participants appear to contribute more in volume online (see Table 3), possibly because the negative effect of age on average contribution length that we see in F2F environments does not occur online. Online environments do not appear to amplify participation inequality related to educational level, which might be a bit of a surprise.

2. *Online effects on individual-level participation equality*. As measured by Gini indices, synchronous voice deliberation (ODP) is on par with F2F methods for individual-level EoP (Table 9). But the optional online Deme forum used in CD produced much greater concentration of participation volume than did F2F methods, including the CD-F2F environment that included the same participants (Table 10).

3. *Online environments and group size effects*. Although the methods tested here are too limited to say so definitively, in this study the online environments (ODP and CD-Forum) eliminated the amplification of inequality that we saw from group size in the BCD and CP (but not CD-F2F) methods.

4. *Online posting as a predictor of F2F participation*. In Table 5 we saw that Forum posters in the CD method out-participated nonposters on all three contribution metrics, indicating that the tendency for an individual to participate is correlated across online and F2F contexts.

5. *Relationship of self-reported experience to measures of participation equality*. The Gini coefficients for frequency, volume, and ACL, as measures of individual-level EoP, proved to be good predictors both of each other and of the subjective equality factor (Table 9, plus Table 11 in [27]). Interestingly, however, at the demographic level there was a more puzzling relationship. Black identified participants rated all but one of the methods more equal than did white participants, even when they participated less by volume than white identified participants did. The ODP method was the only pure test of subjective ratings for an online method, and, consistent with its Gini indices, participants rated it neither the most nor the least equal in comparison to the other (F2F) methods.

For further research, the results related to gender could be taken in a more focused direction. Though ODP was an exercise in synchronous voice deliberation, the purpose of the online forum was question-answering rather than deliberation proper. Using a method similar to CD in which the asynchronous text component were used to deliberate, rather than to share personal anecdotes and ask questions about the topic, would provide a better test of gender equality between online and offline methods. Future research might place more emphasis on individual group composition and its effects on individual contributions, to isolate the cause of demographic trends. Additionally, though outside the scope of this paper, looking at facilitator effects might prove especially useful.